\documentclass[letterpaper,twocolumn,10pt]{article}
\usepackage{hyperref}
\usepackage{usenix-2020-09}
\usepackage[available]{usenixbadges}

\usepackage{amsthm}
\usepackage{graphicx} %
\usepackage{tabularx}
\usepackage{booktabs}
\usepackage{balance}
\usepackage{enumerate}
\usepackage{comment}
\usepackage{url}
\usepackage{tcolorbox}
\usepackage{xcolor}
\definecolor{OxfordBlue}{RGB}{0, 51, 102}

\usepackage{xurl}
\usepackage{xspace}
\usepackage[range-phrase=~--~,range-units=single]{siunitx}

\newtheoremstyle{colonit}%
{}
{}
{\itshape}%
{}%
{\bfseries}%
{:}%
{ }%
{}

\newtheoremstyle{colonrm}%
{}
{}
{}%
{}%
{\bfseries}%
{:}%
{ }%
{}

\theoremstyle{colonit}

\newcommand{\rqtext}{Q}
\newtheorem{rqtheorem}{\rqtext}
\newtheorem{rectheorem}{Recommendation}

\theoremstyle{colonit}

\newcommand{\newcustomtheorem}[2]{%
  \newenvironment{#1}[1]
  {%
   \begin{tcolorbox}[colback=white, colframe=OxfordBlue, rounded corners=southwest, boxrule=0.5mm, title={#2 \ref{##1}}]
   \itshape
  }
  {\end{tcolorbox}}%
  \ifdefined\crefname\crefname{#2}{#2}{#2s}\fi
}
\newcustomtheorem{ratheorem}{Answer to \rqtext}

\DeclareSIUnit{\dB}{dB}
\DeclareSIUnit{\dBm}{dBm}

\begin{document}

\date{}

\title{\Large \bf Current Affairs:\\A Security Measurement Study of CCS EV Charging Deployments}

\author{
{\rm Marcell Szakály}\\
University of Oxford
\and
{\rm Sebastian Köhler}\\
University of Oxford
\and
{\rm Ivan Martinovic}\\
University of Oxford
} %

\maketitle

\begin{textblock*}{\textwidth}(1.9cm,26cm)
\centering
\footnotesize
\noindent
Original Publication by USENIX in Proceedings of the 34th USENIX Security Symposium.
\end{textblock*}

\begin{abstract}

Since its introduction in 2012, the Combined Charging System (CCS) has emerged as the leading technology for EV fast charging in Europe, North America and parts of Asia.
The charging communication of CCS is defined by the ISO 15118 standards, which have been improved over the years.
Most notably, in 2014, important security features such as Transport Layer Security (TLS) and usability enhancements such as Plug and Charge were introduced.

In this paper, we conduct the first measurement study of publicly deployed CCS DC charging stations to capture the state of deployment for different protocol versions and to better understand the attack surface of the EV charging infrastructure.
In our evaluation, we examine 325 chargers manufactured between April 2013 and June 2023, and installed as late as May 2024 by 26 manufacturers across 4 European countries.
We find that only 12\% of the charging stations we analyzed implement TLS at all, leaving all others vulnerable to attacks that have already been demonstrated many years ago.
We observe an increasing trend in support for ISO 15118-2 over the years, reaching 70\% of chargers manufactured in 2023.
We further notice that most chargers use a decade-old firmware for their HomePlug modems, which could contain vulnerabilities that have been patched since.
Finally, we discuss design flaws with the Public Key Infrastructure system used in EV charging, and propose changes to improve the adoption and availability of TLS.

\end{abstract}

\section{Introduction}

\begin{figure}[t]
  \centering
  \includegraphics[width=0.9\linewidth]{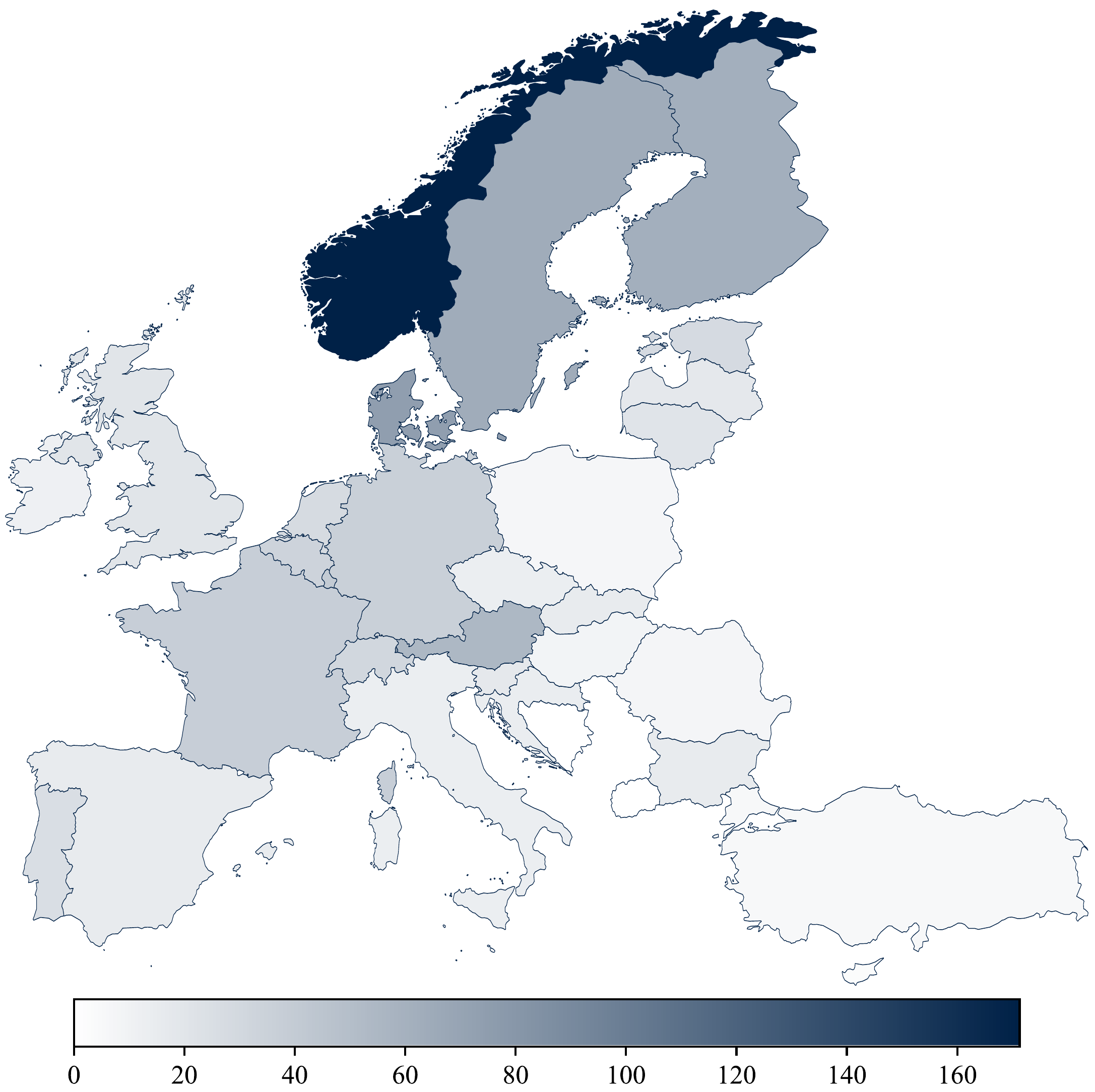}
  \centering
  \caption{DC Chargers in Europe per 100,000 population, based on data from the European Union as of September 2024~\cite{eu_alternative_fuels_observatory}.}
  \label{fig:map_charging_stations} 
\end{figure}

Electric Vehicles (EVs) are rapidly emerging as an important transportation technology, seeing increasing deployment as both personal vehicles and in critical fleets such as logistics~\cite{ev_truck, ev_fleet_amazon}, healthcare~\cite{ev_fleet_nhs}, government~\cite{ev_fleet_us}, mining~\cite{ev_mining}, ferries~\cite{ev_ferry} or public transport~\cite{ev_bus}.
The Combined Charging System (CCS) is the most widely used standard for DC fast charging of vehicles in Europe~\cite{ccs_intro}, and it is also widely deployed in other regions.
CCS is a complicated protocol that offers convenience features in addition to power transfer and uses a modern, digital communication scheme.
This communication carries essential information, including safety-critical power limits.
Researchers have identified a variety of high-impact attacks~\cite{bao2017threat, antoun2020detailed, fries2012electric}, such as Denial of Service~\cite{kohler2023brokenwire}, tampering~\cite{dudek2019v2g}, information theft~\cite{baker2019losing} or electricity theft~\cite{conti2022evexchange}.
They have also shown that despite CCS using a wired communication medium, the physical layer creates an unintentional wireless channel~\cite{baker2019losing, kohler2023brokenwire}, which exposes all of these communication attacks to wireless adversaries.
Modern versions of the standard aim to combat some of these issues by introducing state of the art security, such as TLS 1.3, however older versions remain in use for compatibility.

While TLS addresses many of the security vulnerabilities identified by researchers, we lack information about its actual deployment.
Interoperability is essential for users and regulators to ensure a smooth transition to all-electric vehicles and this requires that each charger and vehicle supports a common version of the protocol.
This is most easily achieved if everyone implements the first and oldest version of CCS as a common fallback.
However, if all chargers and vehicles support a newer version, then old versions can be phased out.
As a result, a key open research question is whether the newer, more secure versions of the standard are being deployed and whether the oldest and insecure 2012 version is being phased out.
In this paper, we address this question with a large-scale experimental study of DC CCS chargers across Europe.

It is often assumed in a variety of domains that insecure and outdated practices remain in use long after better alternatives exist.
However, for EV charging, secure CCS has existed for almost as long as CCS itself, and the field has attracted involvement and standardization from governments.
We believe that regardless of expectations, providing a domain specific snapshot of the industry is important.
To our knowledge, no previous work has examined the deployment of CCS versions.
The most closely related paper~\cite{nasr2023chargeprint} scanned the Internet for exposed management back-ends of EV chargers, whereas our work focuses on the interface between vehicles and chargers.
We summarize our key contributions and findings as follows:

\begin{itemize}
    \item{We designed and implemented an EV emulator, enabling us to collect the first and largest real-world dataset of CCS implementations, comprising data from 325 unique chargers.
    Our design and data will be made publicly available upon publication.}
    \item{Our study reveals that support for Transport Layer Security (TLS) is lacking across all the chargers tested, with only a small fraction (12\%) supporting it.}
    \item{The results also show that ISO 15118-2, a decade-old protocol, has only recently started to gain traction.}
    \item{We discuss the security benefits and implementation cost of TLS.}
    \item{
    During the discussion we identify a trust issue with the current use of TLS certificates.}
    \item{We propose easy-to-implement and backwards-compatible countermeasures to combat this issue, making TLS in CCS more effective and accessible.}
\end{itemize}

\section{Security of CCS}

In this section, we introduce CCS and explain how the protocol works.
Along the way, we identify key security-related design decisions and features, discuss how they are adapted in different versions of the protocol, and pose experimental questions for our study. 

In addition to DC charging, CCS also defines AC charging and bi-directional power transfer using the same protocol.
Furthermore, newer versions of the standard allow for Plug and Charge (PnC), where the vehicle is able to pay automatically.

Besides CCS, other EV charging technologies such as the North American Charging Standard (NACS), formerly known as Tesla Supercharger, CHAdeMO and GB/T exist.
Out of these, the European Union legally requires high-powered DC chargers to offer CCS~\cite{ccs_law}, and the US government only supports charging infrastructure projects that offer CCS~\cite{us_nevi}.
CCS and NACS differ only in the physical connector, but use the same protocols described in the ISO 15118 standards~\cite{std_NACS}.
Due to this wide adoption and standardization, we chose to study public DC CCS charging in this work.
Figure~\ref{fig:map_charging_stations} provides an overview of the current deployment of DC charging stations across Europe.

\subsection{Basic Signaling}

The basic signaling process was developed for use in AC charging and remains a part of CCS for compatibility.
It is a simple Pulse Width Modulated (PWM) signal and does not carry any digital information.
Previous research has demonstrated that this process can be attacked in AC charging by inserting a device into the cable~\cite{zhou2023chargex}, while wireless attacks on PWM signals have also been demonstrated outside the EV context~\cite{dayanikli2022physical}.
However, since CCS uses only a simplified version of the basic signaling for presence detection, these attacks are not applicable to CCS.

\begin{figure}[t!]
  \centering
  \includegraphics[width=0.9\linewidth]{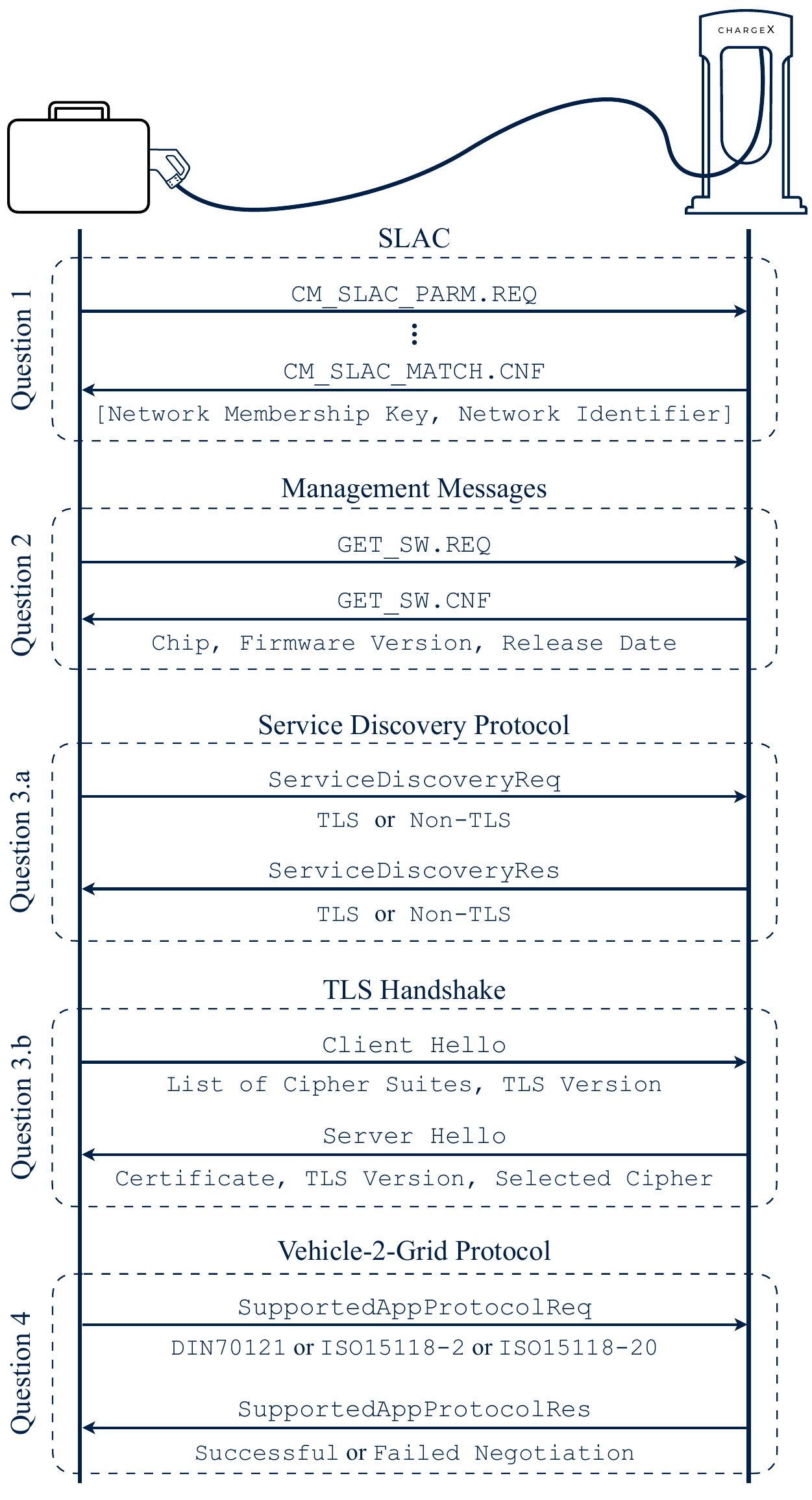}
  \centering
  \caption{The design of our experiments, which follows the same flow as the CCS protocol.
  We show what data is exchanged at each stage in the protocol and how this maps to our experimental questions.}
  \label{fig:ccs_protocol}
\end{figure}

\subsection{Physical Layer}

The ISO 15118-3 \cite{std_ISO_15118_3} standard defines the use of HomePlug Green PHY (HPGP) power line communication as the physical layer for all higher layer protocols.
HPGP is designed to carry Ethernet traffic over just two household mains wires, making it well suited for use in noisy environments over unshielded cables.
To achieve these features, HPGP mimics wireless protocols in structure, using OFDM modulation and robust forward error correction.
However, this RF-like design of the physical layer also allows it to couple into devices either via shared power lines or even wirelessly~\cite{baker2018empower}.
This makes the wired communication channel wirelessly accessible, allowing attackers to intercept~\cite{baker2019losing} and hijack the charging communication in a way that could normally only be possible via a wired Man-in-the-Middle (MitM).
We therefore consider attacks targeting this communication channel to be feasible.%

\subsubsection{SLAC Process}

In HPGP, modems join encrypted networks based on a 128-bit Network Membership Key (NMK), allowing different networks to coexist securely on the same medium.
The NMK is the basis of physical-layer encryption in each network, and knowledge of it allows attackers to join and interact with the traffic flowing through it~\cite{dudek2019v2g}.
Additionally, if attackers have recorded the traffic, they could potentially decrypt it later~\cite{baker2019losing}.
In CCS, the charger and vehicle share the NMK using the Signal Level Attenuation Characterization (SLAC) process.

SLAC serves two purposes: in addition to distributing the NMK, it allows the vehicle to measure which charger it is connected to.
This measurement is necessary when, due to the leakage of HPGP signals, a vehicle can communicate with multiple chargers.
The vehicle measures its attenuation (signal strength) to each charger and selects the best one.
Then, the charger sends the NMK in cleartext to the vehicle.
This allows any attacker present at the start of the charging session to capture the NMK, and access the network~\cite{dudek2019v2g,baker2019losing}.

Alternatively, if an attacker can predict the NMK, they can equally join the network and access the higher level communication at any point during the charging session.
No previous work has studied the existence of weak NMKs, instead capturing the cleartext at the start~\cite{baker2019losing}.
Being able to predict the key and join later during an active charging session would however greatly increase the attack surface and, for example, make drive-by attacks practical.

Each network also has a 54-bit Network ID (NID), which is broadcast in regular unencrypted beacons.
According to the standard~\cite{std_HPGP}, the NID should be calculated deterministically from the NMK using PBKDF1 (a type of hash function) without any additional salt.
In the case of a low entropy NMK, knowing the NID can allow the attacker to determine the NMK via offline computation or a hash table, using attainable resources.
Since the NMK is intended to prevent unauthorized access to the network, a weak or easily guessed NMK could make it possible for an attacker to manipulate the charging communication. 
As a result, we ask the following question:
\begin{rqtheorem}
\label{rq:hpgp_key}
What is the entropy of charger NMKs? Can they be determined from the NID?
\end{rqtheorem}

\subsubsection{HPGP Modem}

Like all complex software, firmware in embedded systems can contain bugs such as buffer overflows and memory corruption.
These can often be turned into code execution attacks, making their potential impact high.
HPGP chips have a complex firmware, handling a large number of management messages, variable length data and networking protocols.
Popular HomePlug chips such as Qualcomm's QCA 7500 have had many public security advisories for vulnerabilities in recent years~\cite{qualcomm_sec}.
Hence, it is important to ensure that HomePlug modems have the latest firmware, leading us to the following question:

\begin{rqtheorem}
\label{rq:hpgp_fw}
What chip and firmware versions do HPGP modems in chargers use? 
\end{rqtheorem}

\subsection{Connection Establishment}
Over the HPGP link, the charger and vehicle communicate using IPv6.
The vehicle begins by sending a multicast Service Discovery Protocol (SDP) request to determine the IP address and port of the charger.
In addition, this process negotiates support for TLS.
The vehicle indicates in the request if it supports TLS, the charger decides what to offer based on the request and its capabilities.
The charger opens the appropriate server socket and sends the IP, port, and TLS support to the vehicle.
If the charger does not offer a TLS connection when asked for one, then it does not implement any form of TLS.
The SDP process is by design vulnerable to a downgrade attack, which could allow attackers to force an unencrypted connection~\cite{zhdanova2022local}.
The simplest mitigation to this is to deprecate non-TLS connections entirely.

During the TLS handshake, the client and server negotiate the supported TLS version, cipher suites, and exchange certificates.
Outdated TLS versions, insecure cipher suites or weak certificates could compromise the security of the TLS connection.
Similar to the NMK, TLS would provide an additional layer of security, making it more difficult for an adversary to eavesdrop on and interfere with the charging communication. 
To understand the attack surface, we pose the following questions: 

\begin{rqtheorem}
\label{rq:tls}
(a) How many chargers support TLS, and have any deprecated non-TLS sessions?
(b) How is it implemented?
\end{rqtheorem}

\subsection{Vehicle-to-Grid}

The Vehicle-to-Grid (V2G) protocol carries the application layer information for CCS, including power delivery negotiation or payment information.
Three versions of this protocol are publicly available, released in three different standards throughout the years.
Newer versions introduce both usability and security features, which we briefly summarize below.%

\textbf{DIN SPEC 70121:}
The first version of the V2G protocol published in 2012, DIN SPEC 70121 contains only the bare minimum needed for power transfer.
It does not use TLS or any other methods to protect the communication.

\textbf{ISO 15118-2:}
Published in 2014, ISO 15118-2 improves on DIN by introducing optional TLS connections, where the charger is the server, and EV is the client.
When TLS is used, it also introduces optional Plug and Charge (PnC), where the vehicle automatically authenticates and pays using a PKI scheme.
It further introduces scheduled charging plans, to take advantage of expected electricity price changes.

\textbf{ISO 15118-20:}
The newest version, ISO 15118-20 introduces new features such as
bi-directional power transfer.
On the security front, it requires mandatory mutual TLS, i.e., the charger and the vehicle both authenticate using certificates~\cite{iso15118_20_features}.

While the standards share many similarities, making it possible to implement them using largely shared code, they are not compatible.
To offer compatibility and allow chargers to implement multiple versions simultaneously, V2G communication starts with a protocol version negotiation.

\begin{rqtheorem}
\label{rq:version}
Which CCS protocol versions do chargers and vehicles support?
\end{rqtheorem}

A simplified, graphical representation of the CCS protocol and its mapping to the questions we will answer in this study is shown in Figure~\ref{fig:ccs_protocol}.

\section{Experimental Methods}

\begin{figure}[t]
  \centering
  \includegraphics[width=0.5\linewidth,trim={0cm 4cm 0cm 0},clip]{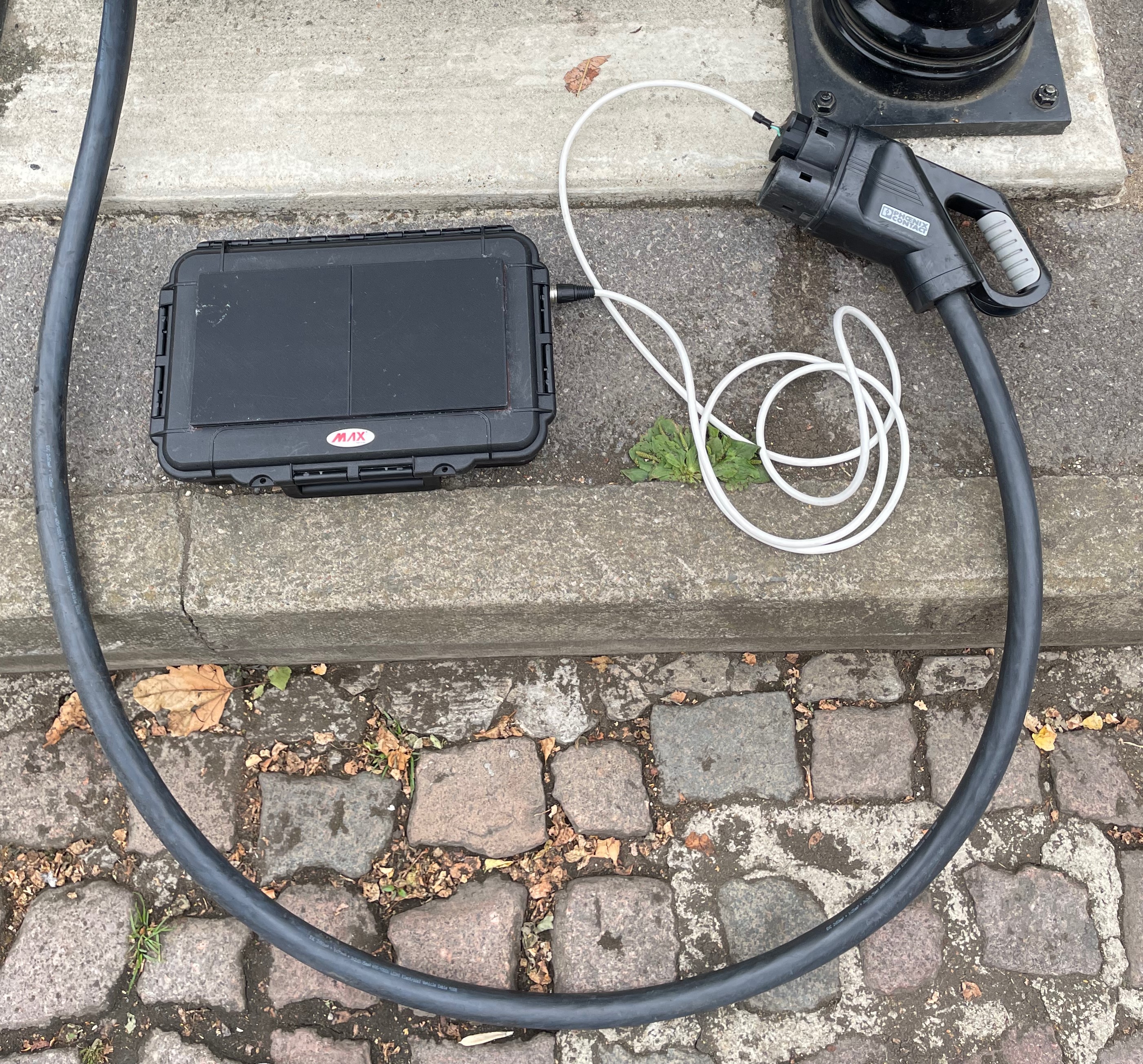}
  \includegraphics[width=0.4\linewidth]{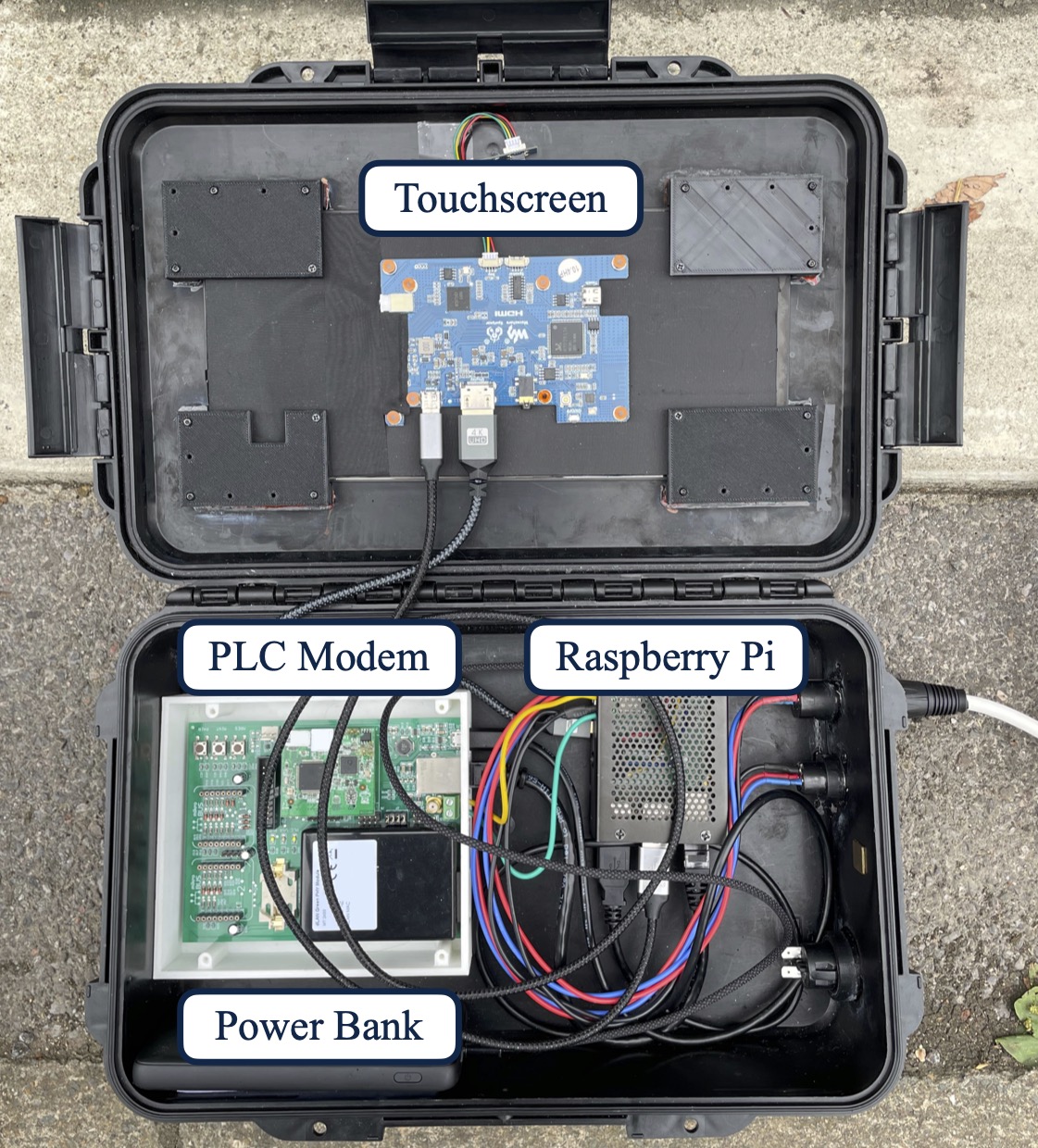}
  \centering
  \caption{The data collection box during one of our experiments. The box contains a touch screen in the top half of the case and power bank, Raspberry Pi and PLC modem in the bottom half.}
  \label{fig:car_simulator}
\end{figure}

Our experimental questions about chargers can be answered by performing multiple CCS charging sessions and studying their behavior.
To achieve this, we developed a vehicle emulator, which implements CCS and saves all relevant information for further analysis.
Unlike a normal vehicle, our emulator implements multiple versions of CCS and alternates between them, thereby querying the charger about its capabilities.
Reading the standards and implementing them ourselves gave us detailed insights into CCS implementations, which aids our discussion later.

\subsection{Emulator Design}

Our vehicle emulator implements the hardware necessary for basic signaling and HPGP, and has the software implementation for the SLAC, service discovery, and V2G communication.
It is built inside of small transport cases as seen in Figure~\ref{fig:car_simulator}, with an integrated powerbank and touchscreen for portability.
Additionally, a web interface on a connected smartphone can control the emulator and provides GPS and camera access for data tagging.

The data collection software runs on a Raspberry Pi 4, extended with a custom versatile PCB that contains the electronics needed to perform basic signaling as an EV.
The Pi is widely available, offers enough performance for a modern graphical operating system and high level programming languages, can easily be powered from a power bank, and has many IO pins for interfacing with custom hardware, which are used by our custom PCB.
To provide HPGP capabilities, a Devolo ``dLAN Green PHY eval board II'' board is connected to the Pi using Ethernet, and configured to EV mode following the instructions in~\cite{open_plc_utils}.
Others have shown that cheap consumer HomePlug AV modems can be converted to CCS-compatible HPGP modems via appropriate configuration~\cite{pyPLC}, but we opted to use a device specifically designed for HPGP EV charging applications.
Our device connects to the charger via a 3D-printed CCS socket, as it would be found on a vehicle.
For safety reasons, we only connected to the data and ground pins necessary for communication and not the power pins.

The Raspberry Pi runs Raspbian, and uses the Linux kernels implementation for the IPv6 and TCP/UDP stacks.
We converted the SLAC and firmware querying tools from the \texttt{open-plc-utils} project by Qualcomm~\cite{open_plc_utils} into a C Python module.
The SDP and V2G protocol stack were written in Python and used a modified version of V2GDecoder~\cite{V2Gdecoder} to translate between the binary and structured representation of V2G messages.%

For the purpose of our evaluation, we extensively logged all valuable data that might help to answer our research questions.
In addition, \texttt{tcpdump} is used to collect full packet captures of all traffic sent over the HPGP network link for manual analysis.

To ensure a correct implementation, we tested our emulator against the \texttt{SwitchEV iso15118}~\cite{switch_iso15118} project, an open source implementation of an ISO 15118-2 and -20 charger, as well as against our own implementation of a charger.
We are thus confident that our implementation is able to collect accurate and meaningful results.
The log files from our script and the raw packet captures are processed automatically into a distilled version of the data.
Additionally, we examined the data for any anomalies that are not part of our pre-existing experimental questions.%

Our implementation differs from real devices in one way: to simplify experiments and facilitate data collection, our circuit board can electrically unplug itself using relays, without the need for a physical disconnection of the plug.
This allows the data collection at each charger to be executed fully automatically, as the emulator can re-connect itself and perform multiple different charging sessions after each other automatically.

\subsection{Charger Measurement Procedure}

To answer our questions about chargers, at each device we perform multiple experiments.
Each experiment consists of a full charging session, as it would be done by a real vehicle, including connecting our emulator, basic signaling, SDP, connection establishment, and protocol negotiation.
After this, the V2G communication is terminated in a proper manner before power delivery, and the emulator electrically unplugs itself, allowing the charger to reset before the next experiment.
Between each experiment, we change the properties of our EV emulator to test for different versions and parts of the protocol.
Some experiments are automatically skipped based on the results of others: for example, if a charger indicates that it does not support TLS, all further experiments that aim to detect the TLS server version are skipped.

To answer \rqtext~\ref{rq:hpgp_key} we collect the NMK and NID from multiple SLAC processes on the same charger.
As each experiment on a charger contains a new SLAC exchange, this data is passively collected.
In addition, to answer \rqtext~\ref{rq:hpgp_fw}, the HPGP device information is queried after every successful SLAC.
To measure TLS support (\rqtext~\ref{rq:tls}), in some experiments we send an SDP query requesting a TLS connection, and in other cases request a non-TLS session.

For chargers that support TLS at the SDP level, we perform further experiments, each using differently configured TLS clients, collecting more data to answer \rqtext~\ref{rq:tls}.
Table~\ref{tab:tls_clients} provides the configuration details for the experiments.
\begin{table}[t]
    \centering
    \caption{TLS configurations for experiments to answer \rqtext~\ref{rq:tls}.}
    \small 
    \begin{tabular}{@{}lll@{}}
        \toprule
        \textbf{TLS} & \textbf{Configuration}\\ \midrule
        1.3 & Mutual authentication for ISO 15118-20 \\ 
        $\geq$1.2 & Only required cipher suites for ISO 15118-2\\ 
        1.2 & Non-standard ``recommended'' cipher suites \\ 
        1.2 & Non-standard and ``not recommended'' cipher suites \\ 
        $\leq$1.1 & - \\ 
        \bottomrule
    \end{tabular}
    \label{tab:tls_clients}
\end{table}
We used the IANA list of TLS cipher suites~\cite{iana_tls} to classify TLS ciphers into ``recommended'' and ``not recommended'' categories.
A ``recommended'' cipher must be widely standardized, considered secure, and offer unique features that other recommended ciphers do not.
Many of the ``not recommended'' ciphers have known security vulnerabilities, including non-ephemeral key exchange, short authentication tags (such as CCM8), and outdated algorithms such as DES.

Finally, with each connection type, we perform experiments to determine the list of supported protocols (\rqtext~\ref{rq:version}).
In all cases, we initiate a V2G session and offer all supported protocols, allowing the charger to indicate the newest version it supports.
In addition, with the TLS 1.3, TLS 1.2 and standard TCP connection types, we perform additional experiments, where each protocol is offered alone in a separate experiment.
This allows us to clearly determine which protocols the charger supports on each connection type.

As we did not have access to a valid ISO 15118-20 vehicle certificate for mutual authentication, we instead generated a self-signed certificate with the correct format.
While it is unlikely that a real -20 charger would accept this, we are still able to infer their support for mutual TLS:
During the handshake, the client (vehicle) does not proactively send its certificate, instead it is sent as a response to a Certificate Request message from the server.
The presence of this request from the server indicates that the charger implements mutual TLS, which is only done for ISO 15118-20.

\subsection{Charging Station Selection}

Due to the large number of EV chargers, we studied a diverse subset of deployments.
To make our data representative of a random subset of devices, we applied a geometric selection strategy, identifying regions or routes with a high charger density, regardless of their manufacturer or network, and attempted to visit all devices in the region.
We noticed that cities, rural areas and highways often host different companies, so we included all of these environments in our dataset, in multiple countries.
We collected enough data to cover most large manufacturers and network operators, with multiple samples in various countries.

In total, we collected data from 149 charging installations (defined as one or more chargers deployed by the same Charge Point Operator (CPO) in close proximity), encompassing 325 chargers and 397 CCS plugs in total.
The chargers were distributed in 4 countries: UK 144 (1.1\%), Switzerland 117 (4.6\%), Croatia 16 (2.9\%), Hungary 48 (6.3\%).
In percentages we show the fraction of the total number of DC chargers in each country as of July 2024\cite{eu_alternative_fuels_observatory} when we conducted our study.
They spanned 26 different manufacturers, with the most plugs tested from ABB (113), Alpitronics (74), Tesla (48), Evtec (25), Efacec (24).
The plugs were located along highways (85), cities (199), and smaller towns (113). %
Devices were manufactured between 2013 and mid 2023, and installed as late as May 2024.
All but one tested charger was manufactured after the public release of ISO 15118-2 and TLS support in 2014, so these features could have been included directly from the manufacturer, without the need for software updates in the field.

\section{Results}

In this section, we present our measurements for each of the key areas investigated and highlight their implications.
Later, the discussion looks at CCS security more generally based on our findings.
When presenting our findings, we report the numbers as seen in our dataset, but later we use information about the known size of various networks to extrapolate our key findings to the wider area.
In rare cases, it happened that an experiment on a particular charger could not be completed while other experiments on the same charger were successful.
For example, we faced time pressure from an EV waiting to charge and could only run the most important experiments.
In such situations, the specific charger was omitted from the particular analysis we did not have data for, resulting in each analysis having a slightly different total number of devices.

\subsection{TLS in Chargers}

\begin{ratheorem}{rq:tls}
All chargers supported non-TLS sessions, 12\% supported TLS, and 6\% implemented standards-compliant TLS 1.2. None implemented mutual TLS.%
\end{ratheorem}

\begin{figure}[t]
  \centering
  \includegraphics[width=0.95\linewidth]{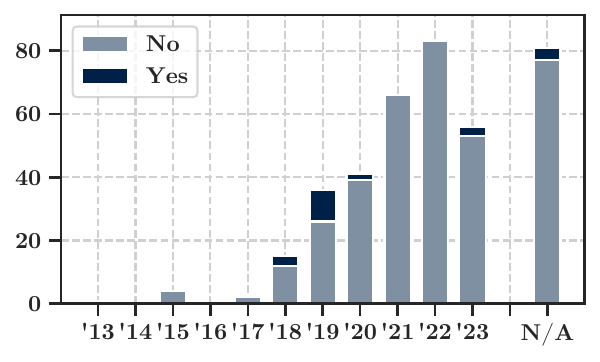}
  \centering
  \caption{Support for TLS according to ISO 15118-2 based on charger manufacturing year in our data. We see no clear correlation with age, but instead a large spike in years when a particular manufacturer actively deployed many new installations across multiple countries.}
  \label{fig:tls2}
\end{figure}

Only \SI{12}{\percent} of chargers from 5 manufacturers in our dataset responded to a TLS SDP request offering a TLS session.
Figure~\ref{fig:tls2} shows the support for ISO 15118-2 compatible TLS, broken down by the manufacturing date of devices, and a full detailed breakdown of all important results is shown in Table~\ref{tab:detailed}.
Our most important observations from the TLS experiments are summarized in Table~\ref{fig:tls2}.

\newcommand{\nwmarker}[1]{#1\xspace}
\newcommand{\nwio}{\nwmarker{Ionity}}
\newcommand{\nwiv}{\nwmarker{Instavolt}}
\newcommand{\nwcp}{\nwmarker{ChargePoint}}
\newcommand{\nwpo}{\nwmarker{Porsche}}

\newcommand{\mfmarker}[1]{#1\xspace}
\newcommand{\mfal}[0]{\mfmarker{Alpitronics}}
\newcommand{\mfek}[0]{\mfmarker{Ekoenergetyka}}
\newcommand{\mfcp}[0]{\mfmarker{ChargePoint}}
\newcommand{\mfpo}[0]{\mfmarker{Porsche}}
\newcommand{\mfef}[0]{\mfmarker{Efacec}}

\subsubsection{Legacy Implementation}
In our testing, one manufacturer used by two networks in multiple countries, offered TLS on all their tested devices, however they all refused connections from a TLS 1.2 client configured according to the standard.
The standard requires using elliptic curve key exchange and certificates, instead these devices accepted TLS 1.2 and TLS 1.1 connections using an RSA based key exchange.
They all had the exact same RSA-1024 certificate, that expired in 2013, long before their manufacturing date.
We assume that this was implemented as part of testing during the standardization of ISO 15118, and was not removed from production devices.
In addition, they do not advertise Plug and Charge capabilities.
For further analysis we do not count these devices as supporting TLS, as they are not compatible with standard compliant vehicles expecting elliptic curves or up-to-date certificates.
Accordingly, their security is equivalent to that of a charger with no TLS support.

\subsubsection{Certificates}

Considering only the implementations of TLS 1.2 and above, we examined the charger certificates, which could be traced back to root certificates from two of the largest V2G PKI providers, Hubject or Entrust.
Both PKI providers implement the certificates correctly, following the certificate types and structure described in the standard.
All of the European PnC networks used a Hubject root certificate.

The Entrust certificate was used by all tested devices from one of the manufacturers, deployed by 2 different networks.
Furthermore, these devices shared one of only two leaf certificates, with a precisely 1 year shift in their window of validity.
The certificates appeared to not be updated, since one of the two certificates had expired 3 months prior to testing.
We were already in contact with this company due to a vulnerability explained later, and also reported this.
They confirmed that the devices were not meant to have TLS enabled, and were shipped with a development version and certificates.
This is further confirmed since neither the manufacturer, nor the two networks using them currently advertise Plug and Charge capabilities, and the certificate content indicated that they were meant for the US market, using the US Entrust root.
Our assumption is further backed up by observations of one of the two networks, who only had TLS enabled on their chargers from this manufacturer, but not on devices from other manufacturers.

\subsubsection{Version Support}

ISO 15118-2 requires the use of TLS 1.2 or higher, while -20 requires mutual TLS 1.3.
Older versions should not be used, and cipher suites should be limited to those listed in the standard.
In our dataset we found no chargers performing mutual TLS, and only \mfcp had TLS 1.3 enabled.
TLS 1.2 was implemented on devices by 4 manufacturers (\mfal, \mfek, \mfcp, \mfpo) supplying 4 networks (\nwio, \nwiv, \nwcp, \nwpo), of which 2 networks (\nwio, \nwpo) advertised Plug and Charge capability.
Out of these 4 manufactures, one implemented unnecessary ciphers, however this was an IANA recommended cipher, and thus safe by current standards.
However, this still highlights that real-world implementations can contain misconfigurations and deviations from the standard, exposing a larger attack surface than necessary.
Only the company who used RSA certificates implemented not recommended ciphers.

Out of the Plug and Charge compatible networks, \nwio was supplied by two manufacturers (\mfal, \mfek), but implemented TLS and Plug and Charge in both cases.
Finally, \nwpo uses only their own custom chargers and also implemented TLS consistently.

\begin{figure}[t]
  \centering
  \includegraphics[width=0.95\linewidth]{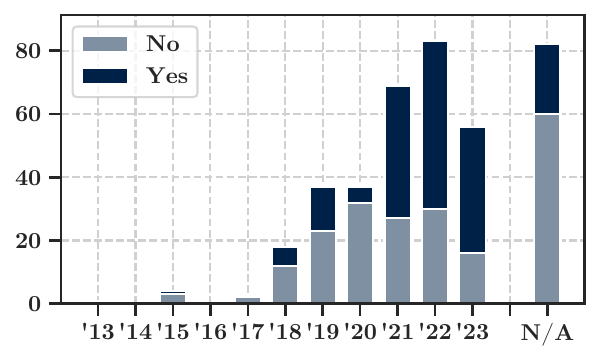}
  \centering
  \caption{Support for ISO 15118-2 based on charger manufacturing year in our data. We see that new devices increasingly deploy this new version.}
  \label{fig:iso2}
\end{figure}

\begin{table*}[t]
    \centering
    \caption{Detailed results for the key experimental questions, broken down per manufacturing year. Each cell contains the number of plugs as Yes/No. Note that not all experiments produced useful data, so the total number of experiments in each case may be different.}
    \label{tab:detailed}
    \resizebox{1\textwidth}{!}{%
    \begin{tabular}{lccccccccccccc}
        \toprule
        & & \multicolumn{12}{c}{Manufacturing Year}\\
        \cmidrule(lr){3-14}
          & All & Unknown & 2013 & 2014 & 2015 & 2016 & 2017 & 2018 & 2019 & 2020 & 2021 & 2022 & 2023\\
        \midrule
        TLS in SDP & 47 / 344 & 7 / 76 & 0 / 1 & 0 / 0 & 0 / 4 & 0 / 0 & 0 / 2 & 12 / 6 & 18 / 19 & 5 / 36 & 0 / 66 & 2 / 81 & 3 / 53 \\
        TLS 1.3 for ISO 15118-20 & 4 / 380 & 4 / 77 & 0 / 1 & 0 / 0 & 0 / 4 & 0 / 0 & 0 / 2 & 0 / 14 & 0 / 36 & 0 / 41 & 0 / 66 & 0 / 83 & 0 / 56 \\
        TLS 1.2+ for ISO 15118-2 & 22 / 363 & 4 / 77 & 0 / 1 & 0 / 0 & 0 / 4 & 0 / 0 & 0 / 2 & 3 / 12 & 10 / 26 & 2 / 39 & 0 / 66 & 0 / 83 & 3 / 53 \\
        TLS 1.2 recommended ciphers & 3 / 371 & 0 / 76 & 0 / 1 & 0 / 0 & 0 / 4 & 0 / 0 & 0 / 2 & 0 / 12 & 0 / 33 & 0 / 41 & 0 / 66 & 0 / 83 & 3 / 53 \\
        TLS 1.2 not recommended ciphers & 3 / 371 & 0 / 76 & 0 / 1 & 0 / 0 & 0 / 4 & 0 / 0 & 0 / 2 & 0 / 12 & 0 / 33 & 0 / 41 & 0 / 66 & 0 / 83 & 3 / 53 \\
        TLS 1.1 and older & 14 / 369 & 0 / 80 & 0 / 1 & 0 / 0 & 0 / 4 & 0 / 0 & 0 / 2 & 6 / 8 & 6 / 30 & 2 / 39 & 0 / 66 & 0 / 83 & 0 / 56 \\
        \midrule
        ISO 15118-20 DC & 0 / 380 & 0 / 80 & 0 / 1 & 0 / 0 & 0 / 2 & 0 / 0 & 0 / 2 & 0 / 17 & 0 / 37 & 0 / 36 & 0 / 68 & 0 / 83 & 0 / 54 \\
        ISO 15118-2:2013 & 183 / 206 & 22 / 60 & 0 / 1 & 0 / 0 & 1 / 3 & 0 / 0 & 0 / 2 & 6 / 12 & 14 / 23 & 5 / 32 & 42 / 27 & 53 / 30 & 40 / 16 \\
        ISO 15118-2:2010 & 33 / 348 & 2 / 79 & 0 / 1 & 0 / 0 & 0 / 2 & 0 / 0 & 0 / 2 & 0 / 17 & 1 / 36 & 0 / 36 & 1 / 67 & 20 / 63 & 9 / 45 \\
        DIN SPEC 70121 & 389 / 0 & 81 / 0 & 1 / 0 & 0 / 0 & 4 / 0 & 0 / 0 & 2 / 0 & 18 / 0 & 37 / 0 & 39 / 0 & 69 / 0 & 82 / 0 & 56 / 0 \\

        \bottomrule
    \end{tabular}%
    }
\end{table*}

\subsubsection{Implementation Strategy}
One of the largest manufacturers, \mfal, supplies many different network operators.
Their \nwio chargers use TLS and PnC, but none of the others offer PnC or have TLS enabled.
Based on these observations, we assume that the TLS server software is implemented by the manufacturer, but the network operator decides whether to enable the feature and provides the certificates to do so.
In the case of the company that used RSA or shared Entrust certificates, they released a development firmware with TLS enabled using development certificates into public devices, as we observed their devices exhibiting the same behavior on different networks with the same certificates.
In networks where TLS and PnC are intentionally deployed (\nwio, \nwpo), the certificates are supplied by the network operator.

\subsubsection{Additional Observations}

One manufacturer did not implement RFC 5746 \cite{rfc5746} from 2010, which signals mitigation for the CVE-2009-3555 TLS renegotiation vulnerability.
This was a vulnerability in the TLS protocol, and alongside fixing the issue, a new extension was introduced to TLS, allowing clients and servers to signal that they implement the revised version of the protocol.
To our knowledge, both the mitigation and this signaling are implemented and enabled by default in all major TLS libraries, indicating either that it was manually disabled, or that they are using a very outdated library.
The vulnerability allows any attacker to inject data into the start of a TLS session.
When TLS is used as part of HTTPS, this exposes a powerful attack against the HTTP protocol.
However, we are not aware of a way this could be usefully exploited against the binary encoding of the V2G protocol.
Also, for ethical reasons we did not actively exploit this vulnerability, so it remains unknown if the device is vulnerable to the underlying attack.
As per their vulnerability disclosure program, ``Missing best practices in SSL/TLS configuration'' and ``Previously known vulnerable libraries without a working Proof of Concept'' are out of scope.

One manufacturer's chargers sent cleartext UDP packets to the EV during the TLS handshake.
These packets contained information in the NSS Key Log format~\cite{nss_keylog}, and when their contents are loaded into standard tools such as Wireshark, they allow the TLS connection to be decrypted~\cite{wireshark_sslkeylog}.
A similar behavior is described in~\cite{mahadevegowda2022secure}, who explain that sending key info in UDP packets exactly as we observed is used in debugging tools by automotive vendors to decrypt and debug the TLS connections.
However, this could similarly be performed by any attacker who has access to the physical layer, defeating the Diffie-Hellman key exchange in TLS.
We reported this via email to a contact at the manufacturer in July 2024, who confirmed that it was intended as a debugging feature that accidentally made it into production.

\subsection{Standard Support}

In our experiments, chargers were offered all known CCS protocol versions individually.
This allowed us to  unambiguously determine which protocols a device supports, based on whether it accepts the protocol.
Our results about supported protocols in chargers can be seen in detail in Table~\ref{tab:detailed}.
\begin{ratheorem}{rq:version}
All chargers support DIN SPEC 70121, and 47\% implement ISO 15118-2. No chargers implemented 15118-20.
All 8 tested vehicles supported DIN, and 7 supported 15118-2.
\end{ratheorem}

Despite ISO 15118-20 capable cars being advertised and sold, we are not aware of any public charger advertising this capability.
Our testing included PnC capable chargers deployed at dealerships where ISO 15118-20 capable vehicles are sold, and even in this case we did not detect support for TLS 1.3, mutual TLS or ISO 15118-20.

Much like how all chargers offer non-TLS connections for compatibility, as the oldest and simplest version, DIN is still widely used as the common fallback option.
However, an important trend visible in the data is a slow rise in ISO 15118-2 support based on manufacturing year, as shown in Figure~\ref{fig:iso2}.
While most of these devices offered only the insecure version of ISO 15118-2, it is nonetheless important for newer versions of the protocol to be implemented.

\subsection{HPGP Firmware}

\begin{figure}[t]
  \centering
  \includegraphics[width=0.9\linewidth]{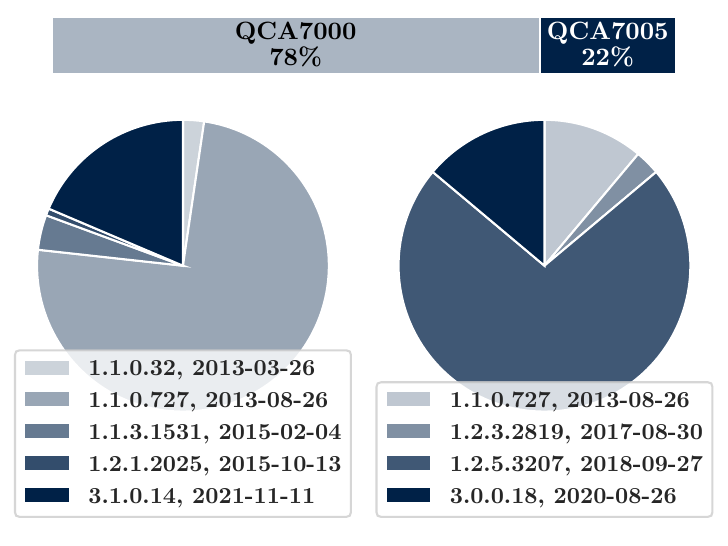}
  \centering
  \caption{Firmware versions used by the QCA 7000 (left) and QCA 7005 (right) chips deployed in EV chargers.}
  \label{fig:firmware}
\end{figure}

\begin{ratheorem}{rq:hpgp_fw}
All chargers used Qualcomm HPGP chips, and 62\% had more than 10 years old firmware.
\end{ratheorem}
The breakdown of chip type and firmware versions can be seen in Figure~\ref{fig:firmware}.
In all of our testing, devices used one of two different HPGP chips, the QCA 7000 (78\%) and QCA 7005 (22\%).
Based on available information these chips appear to be nearly identical: they are advertised as pin compatible, share the same internal design and even firmware.
The only publicly known difference between them is their temperature rating and release date.
The most common firmware versions across a wide variety of manufacturers was 1.1.0.727 from 2013 for the QCA 7000 and 1.2.5.3207 from 2018 for the QCA 7005.
However, the newest versions seen were 3.1.0.14 from 2021 and 3.0.0.18 from 2020 respectively, showing that new revisions are still being developed.
In one instance, a charger manufactured in 2020 was using firmware from 2021, indicating either that it had a hardware replacement, or that the HPGP firmware can be updated.

Qualcomm does not provide much public information about these chips, or their firmware, since they are not intended directly for end users.
As such, we do not have access to the change logs and cannot assess the risk of outdated firmware.

According to the Qualcomm Security Bulletin~\cite{qualcomm_sec}, there are no public vulnerabilities in the QCA 7000/7005.
However, we were able to find entries for many different chips, including the QCA 7500 chip, a HomePlug AV2 (HPAV2) modem.
Therefore, we believe it is likely for the QCA 7000 firmware to also contain vulnerabilities, including some that may have been discovered or patched.
A culture of up-to-date firmware will help prevent issues when they are discovered.

Brokenwire is a known attack against HPGP, which wirelessly performs a highly efficient jamming and Denial of Service (DoS) against the wired communication~\cite{kohler2023brokenwire}.
While jamming is an unfixable issue, Brokenwire combines the fact that HPGP modems are extremely sensitive to a weak packet preamble, and that they use Carrier Sense Multiple Access with Collision Avoidance.
This means that an attacker repeatedly transmitting a preamble at the same power as the noise floor will cause transmitters to stop transmitting entirely, causing the CCS charger to perform an emergency stop.
This behavior should be controlled by firmware, making it possible for Brokenwire to largely be fixed by software.
According to the ethics of our work, we did not test for the attack, however no charger used firmware released after the disclosure of the attack in 2022, so it must be currently unmitigated.
Additionally, the outdated state of firmware indicates that even if a patch were released, it will not be widely deployed for a long time.

\subsection{Encryption Keys}

\begin{ratheorem}{rq:hpgp_key}
Most chargers used random, ephemeral NMKs and standard compliant NIDs, with a few exceptions.
\end{ratheorem}

In line with the standard~\cite{std_ISO_15118_3}, we found that in most cases a new, random NMK is used for each vehicle connecting, and the NID is derived from the NMK using PBKDF1, as specified by~\cite{std_HPGP}.
Assuming the random seeming bytes are generated by a high quality random generator, these NMKs have 128 bits of entropy.
This cannot be feasibly cracked, even using the information from the 54 bit NID.

Our analysis revealed 2 manufacturers where the NMK contained 12 ephemeral random bytes, as well as 4 fixed bytes, and the NID was still calculated using PBKDF1.
We do not know why reducing the key space is practical, particularly when the random generator has already been implemented for the remaining bytes.
This reduced 96 bit entropy could make cracking the key easier, but it is still sufficiently secure.
As no salt is used, and the fixed bytes are shared between many chargers of the same model, hash tables may be practical to accelerate the attack.

Additionally, one manufacturer exhibited a very unique behavior, where many of their NMKs contained long sections of previous keys, often shifted by one byte.
This behavior could allow attackers who captured a single NMK to determine subsequent ones in a greatly reduced search space.

Finally, we found one manufacturer who did not use PBKDF1 to derive the NID from the NMK.
Instead, they generated 4 random bytes, which were then padded with a fixed pattern to get both the NID and NMK.
Not only does this NMK only have 32 bits of entropy, it can trivially be derived from the NID.

In all cases, the NMK only protects against attackers who are not present at the start of the charging session.
By design, the NMK is sent in clear text during the initial SLAC process and can be easily captured by wireless attackers~\cite{baker2019losing}.

\section{Discussion}

The charging communications carry safety critical information about the battery voltage and capabilities of the vehicle.
However, an attacker with access to the communications could modify these values and cause the battery to over-charge.
A well designed vehicle should perform an emergency disconnect when it detects this, but this requires a physical disconnection of contactors while under a very high current (up to \SI{500}{\ampere} is used in DC charging), which could easily damage them due to arcing.
Previous work~\cite{dudek2019v2g} has shown that MitM access to V2G communication is possible, and it is also known that wireless access is feasible~\cite{baker2019losing}.
Therefore, it is essential to secure CCS communication against tampering.

Without major changes to the standard, there are two avenues for protecting the communications.
Firstly, the HPGP network could be protected by improving how the NMK is distributed, thereby securing all other parts of the protocol.
No currently existing standard revision addresses this issue, but we highlight a potential avenue below.
Secondly, the TCP link that carries the data could be upgraded to TLS.
This method is not new, as optional TLS was introduced in 2014 by the ISO 15118-2 standard.
Our results show a rising trend in support for 15118-2, but in 2024 it is still common to see newly deployed chargers that only support DIN 70121, and most devices do not support TLS at all.

\begin{rectheorem}
    All chargers and vehicles should be upgraded to support at least ISO 15118-2 with TLS.
    Government regulation or subsidies can be used to encourage companies to update quickly.
    Once a sufficient threshold is reached, insecure communications should be deprecated.
\end{rectheorem}

In addition to the clear security benefits, ISO 15118-2 and TLS also enable use cases beneficial to consumers:

\textbf{PnC} allows EVs to authenticate and pay autonomously, allowing owners to charge without interacting with an app or payment terminal.
This user experience is so desirable that the Autocharge system~\cite{autocharge} was developed to offer the same capabilities with simpler technology on the older standards.
It continues to be used, despite significant known security issues that can allow attackers to easily impersonate victims and steal electricity from them~\cite{baker2019losing}.
Some companies refuse to implement Autocharge because of these security issues~\cite{autocharge}.

PnC provides the same user experience with added security, but based on our observations it is not currently widely used.
It will take additional work to implement the PnC protocol on the front and back ends, but this should be comparable to other payment solutions that are already in use.
Open source reference implementations exist, and additionally the Open Charge Point Protocol (OCPP) describes a standardized and interoperable system for the necessary back-end communications.
In addition, by design, PnC requires collaboration between EV manufacturers, charger operators, and payment providers.
This also requires negotiations and business relations between the relevant companies.

\textbf{Bidirectional and Scheduled Charging} are important new features for the future of the power grid, as they allow vehicles to charge at times of high supply, or to feed energy back into the grid during peak demand.
Such features can be particularly useful for owners of home photovoltaic installations, or large vehicle fleets, who can optimize their energy usage.
In addition, a distributed network of EVs plugged in and acting as battery buffers are often considered to be an important element of a stable, renewable energy grid.
Charging station manufacturers often make devices for small home and office installations, private fleets, and public DC fast charging.
So, even if some of these features might not make sense in a fast charging environment, manufacturers have a clear incentive to implement them for their other customers.
These drive the implementation of the underlying new standards, which can then be deployed on all their devices.

\subsection{Implementation}

With these benefits and motivation in mind, we consider the cost of implementing TLS and modern protocol versions.
Implementation of new standards requires additional engineering work, since each version has incompatible data structures and binary encoding.
However, the technologies involved are largely similar, allowing manufacturers to easily update.
All hardware aspects, including basic signaling and HPGP are unchanged between CCS standard versions, ensuring that no hardware changes are necessary.
Additionally, SLAC, and SDP are also identical.
The difference is in the V2G messages, and their binary encoding.
These are clearly documented in the standard, and multiple open source implementations can serve as reference~\cite{everest, switch_iso15118}.
Further, in our experience, DIN and ISO 15118-2 are so similar that they can be implemented with essentially the same code, containing only minor differences.
Consequently, adding ISO 15118-2 to a charger that already implements DIN is only a small fraction of the original effort needed to implement it.

To implement TLS, simple, free, and extensively tested libraries exist for all major programming languages.
Modern micro-controllers should easily handle the additional calculations, and PnC also does not require new hardware.
Despite this, for some car models~\cite{pnc_bmw}, support for PnC depends on the date of manufacturing.
We assume that this is because manufacturers are required to use Hardware Security Modules (HSMs) to store private keys~\cite{mahadevegowda2022secure, hubject_pnc_whitepaper}, which are not available in older hardware.

\begin{rectheorem}
All devices should start shipping with the necessary TLS hardware as soon as possible, even if the software deployment is delayed.
\end{rectheorem}

\subsection{Public Key Infrastructure}

In addition to the simple implementation work, TLS requires a functioning Public Key Infrastructure (PKI) to provide roots of trust.
The standard~\cite{std_ISO_15118_2} does not define a central Certificate Authority (CA), but in the notes it describes a vision where each continent has no more than one dominant operator.
This would allow all manufactured vehicles to ship with a list of trusted certificates, much like how browsers and operating systems come pre-installed with trusted root CAs for the web.

In our measurements, we conclude that the market has converged on a single dominant operator in Europe, as we have found that all PnC networks use the same Hubject root certificate.
Additionally, we saw evidence of an Entrust root certificate meant for the US market in some of our tests.
We only found this in the case where devices were shipped with a development TLS server enabled, as discussed earlier.
The limited number of root certificates allows vehicles to easily include a comprehensive list from the factory, allowing them to validate chargers, regardless of PnC membership.

Based on their product page, Hubject currently charges a one time fee of €3900 per company, and a further subscription fee of €0.99 per charger per year for access to a PnC certificate~\cite{hubject_cert_price}.
We think that both of these fees are well within the budget of an EV charger company, and therefore should not discourage the implementation of TLS.

Based on the results of our study, we believe that TLS is being treated as a necessary condition to enable PnC, instead of as a security feature for power delivery messages.
In our work, we did see chargers that deployed TLS without PnC, however as described in our results section, these appears to be development tests instead of real deployments.

\subsection{TLS Design Issue}

During our work analyzing the results of the study, we identified a previously unknown flaw with the current PKI system.
This flaw could allow attackers with a single valid TLS certificate for any charger to attack all communications using the same root CA.
To explain this issue, we highlight a key difference between TLS as it is used on web, and in CCS.

In both cases, a TLS certificate needs to be signed by a root CA that is trusted by a client.
These roots come pre-installed in devices, or can be added by the user.
A website or charger owner can approach a CA, and request a certificate for their website or charger.
The CA verifies the identity or ownership necessary for the request, and the EVSE ID or the website URL are encoded into the issued certificate.
When a browser connects to a specific website \texttt{{good}.example}, but an attacker performs a MitM attack and uses their certificate for \texttt{{evil}.example}, this will be rejected by the browser.

Similarly, the charger certificates contain the EVSE ID, which is unique to each charger.
However, the same check cannot be performed by the car, as it does not know which charger it is expecting to connect to.

Any charger certificate issued to any charger by any trusted root CA will be accepted in any location.
There are many ways a certificate could be compromised, e.g., via extracting it from a charger in the wild, or by an attacker working in the supply chain.
With this certificate, the attacker can MitM all TLS protected charging sessions globally for that CA, until their certificate expires or is revoked.
To address this, the CAs must strictly evaluate any entity they issue certificates to, increasing the cost and difficulty of implementing TLS significantly.

We propose two avenues to increase the resistance of the PKI system against such an attack, greatly restricting the capabilities of an attacker with a stolen certificate.

\paragraph{Location Restriction}
A vehicle should be able to check if the certificate presented belongs to the charger it is connected to, using only information it can obtain from a trusted source.
Therefore, we propose to add the GPS coordinates of the charger into the certificate, which could be done by using the Subject Locality attribute, which is not currently used.
Since most vehicles nowadays contain GPS, they could easily verify that their location is close to the chargers claimed location.
When issuing certificates, CAs could require proof that a charger is being installed at a specific location, and monitor for overlapping requests or an entity requesting suspiciously high numbers of new locations.
With this method, the impact of a single attacker controlled certificate will be limited to a small area.

\paragraph{OCSP Stapling}
If a compromised certificate is discovered, it is essential to quickly revoke it.
The most used solutions to this are the Online Certificate Status Protocol (OCSP), as well as Certificate Revocation Lists (CRLs)~\cite{chung2018web}.
OCSP provides a method for a client to query the CA about the current validity of a certificate.
If the certificate needs to be revoked, the CA can decline all further queries for validity.
CRLs are lists published by CAs, listing all certificates they have issued, which have been revoked but not yet expired.
Both of these methods require the vehicle to access the internet to contact the CA.
While many modern vehicles contain mobile data connection, this may not always be available.
Additionally, there are privacy concerns associated with OCSP, as it could allow the CA to track clients.
OCSP Stapling~\cite{rfc6960} addresses both these issues.

In OCSP stapling, the TLS server sends the results of a recent OCSP query to the client, along with their certificate.
The ISO 15118-2 standard requires all chargers to provide OCSP responses in their TLS handshake for Sub-CA certificates in the chain, but not the leaf~\cite{std_ISO_15118_2}[V2G2-070, Note 1].
Vehicles are also not required to verify the leaf by other means.

In our measurements, the chargers did not provide an OCSP value for the leaf.
As the attacker is most likely to compromise the leaf certificate, we believe this this is an important safety feature, which can be introduced retroactively.

\begin{rectheorem}
    Chargers should add OCSP responses for the leaf certificate, and vehicles should validate the leaf.
\end{rectheorem}

\subsection{Secure SLAC}

Finally, we discuss a method to protect the communication at a layer below TLS.
While TLS provides confidentiality and authentication of the V2G data, an attacker who is able to access the HPGP network is presented with a large attack surface, including the HPGP modems, Ethernet interface of the car and charger, SDP process, or TLS implementation flaws.
A simple solution to all of these issues is to prevent third parties from interacting with the network, thereby adding an additional layer of defense.
HPGP provides full protection of the network from anyone who does not have access to the NMK.
However, in the SLAC process currently used, the NMK is distributed in clear text visible to any attacker.
The HPGP standard defines the Secure SLAC process~\cite{std_HPGP}, where the SLAC process (and key exchange) are cryptographically protected by PKI based cryptography.
However, all versions of CCS explicitly specify using only insecure SLAC.
To our knowledge, no implementation of secure SLAC exists, nor has it been studied by researchers.
The secure SLAC key exchange process encrypts and authenticates the NMK using public key cryptography, therefore our previous discussion of PKI applies to this protocol as well.
Future versions of the standard should modify the SLAC process to protect the NMK, such as by implementing a combination of Secure SLAC, and the Diffie-Hellman protocol proposed in~\cite{baker2019losing}.

\subsection{Future Outlook}

At the time of our study, there were approximately 147,208 DC chargers across Europe~\cite{eu_alternative_fuels_observatory} (data from September 2024).
This number is rapidly increasing, and their deployment seems to be accelerating.
Therefore, it is essential that newly deployed devices come equipped with all the necessary hardware and potentially software directly from the factory, in order to avoid costly in-field upgrades.

In addition to this, a paradigm shift in AC charging will soon majorly boost the deployment of CCS.
Currently, most AC chargers are very simple devices, acting as an electricity meter and switch between the electrical supply and the charging cable.
For safety reasons, basic signaling is used for presence detection of the vehicle before powering the cable.
However, the CCS standard defines using the same sophisticated communication scheme for AC charging in addition to DC.
This allows users to schedule charging based on varying electricity prices, and use PnC.

As of our data from September 2024, 84\% of all chargers in Europe are AC~\cite{eu_alternative_fuels_observatory}, and we believe that most of these do not yet use CCS.
As ISO 15118 capable AC chargers are adopted, they will rapidly outnumber CCS DC chargers.
The security implications of AC charging communication attacks must be studied further, but attackers could increase the current drawn by vehicles, potentially damaging the cable or tripping fuses.
Additionally, they could tamper with scheduled charging, denying service or inflating electricity costs.

Our work focused on chargers, however we also briefly investigated the protocol deployment in EVs.
Based on an online dataset of EV capabilities discussed in Appendix~\ref{app:ev}, newer and premium vehicles are the first to adopt TLS support and most vehicles that have been sold do not support TLS.
In some cases, even within the same model only newer vehicles support PnC, while older ones do not~\cite{pnc_bmw}.
In order to phase out non-TLS protected charging sessions, both vehicles and chargers will need to support the feature.

\section{Limitations and Extrapolation}

We acknowledge that our survey provides only a snapshot of the current state of EV charging infrastructure deployment across Europe. 
Our study included countries with mature and developing EV networks, as well as diverse locations such as dense urban cores, rural deployments, and large highway infrastructure projects.
Because of this, we believe that our key findings are representative of the state of EV charging in Europe, as of early September 2024.
This assertion is further supported by the widespread presence of many network operators in different countries, and their use of the same charging station manufacturers.

We observed that the chargers deployed by a given network and manufactured by a specific company exhibited consistent behavior regardless of location.
Therefore, based only on information about the network and manufacturer of each charger, we can make predictions about their behavior.
Motivated by this observation, and the existence of large international networks, we are able to extrapolate our existing measurements, and identify the portion of the market not covered by our experiments.
We used Zap-Map~\cite{zap-map}, a large database of EV chargers across Europe.
As a UK based company, they provide detailed information about the UK, so we chose to extrapolate our findings to this country.
From this data source, we obtained information about the size of various charging networks.

To estimate the usage of various manufacturers by a given network, we used public photos of a random subset of their chargers and identified the device based on distinctive visual features.
For some networks, it was even possible to identify the exact model of the charging stations based on the unique id, which in addition to the manufacturer included the exact model.

\subsection{TLS Support}

Using this collected data, we counted the TLS supporting networks we identified.
In total, we found that \SI{3.1}{\percent} of the chargers are from networks and manufacturers where we expect TLS support.
Similarly, we estimate that \SI{74.5}{\percent} of chargers are from networks that do not support TLS.
The remaining devices are from a combination of many networks which did not appear in our dataset.
Based on our observation that TLS is deployed either by networks with Plug and Charge, or by devices from a manufacturer who ships a TLS server with development certificates, we assume that the remaining chargers are unlikely to support TLS.

\section{Related Work}

Survey studies are commonly done in many different areas, to allow future researchers and the public to understand the state of affairs.
In the security context, non-malicious internet scans are regularly performed by researchers~\cite{wang2022large, roomi2023large} and cyber security companies~\cite{shadowserver, shodan} to understand the status of public facing infrastructure, or to analyze long term trends.
In particular, the configuration of TLS servers on the internet has been measured over multiple years to analyze trends~\cite{kotzias2018coming}.
In the EV context, researchers have scanned for web interfaces of chargers intended for home use~\cite{nasr2023chargeprint}, as well as for endpoints of the Open Charge Point Protocol (OCPP)~\cite{sarieddine2024uncovering}.
In both cases, they identified a large number of public devices, and discovered new vulnerabilities in them.
Non-security related EV charging studies have also been conducted.
In~\cite{chen2020review}, the authors present a review of EV charging in the UK, with a focus on geographic distribution, power usage, and the planning process.
In addition, multiple studies offer a comprehensive overview of EV charging, discussing non-CCS technologies, capabilities of EVs, legislation and market trends~\cite{mastoi2022analysis, amel2024charging}.
To our knowledge, this paper is the first project performing a study of chargers via the CCS interface, instead of via the network, complementing the existing work.

\subsection{Implementations}

Similar to our EV emulator, various open source projects~\cite{pyPLC, switch_iso15118, rise_v2g, everest} implement the communication stack of CCS, including power delivery messages.
Some of them feature implementations of Plug and Charge, and act as valuable research tools for those wishing to study this protocol in action.

\subsection{Attacks}

Due to its importance, many papers examine the security of CCS and EV charging in general.
Overview papers such as~\cite{bao2017threat} enumerate the possible motivations and entry points for attackers.

Researchers have demonstrated the extraction of the NMK from the SLAC~\cite{dudek2019v2g}, even wirelessly~\cite{baker2019losing}.
Dudek et al.~\cite{dudek2019v2g} further demonstrated that they can join the network and inject packets into the communication.
A TLS based session secures against passive eavesdroppers and should even remain secure against most forms of active MitM attacker, as long as they do not possess a trusted certificate belonging to a charger.
As we discussed, a single compromised certificate could enable TLS attacks globally, however it is still increases the bar for attackers, and therefore TLS should be enabled.
Physical layer wireless jamming such as~\cite{kohler2023brokenwire} could potentially be counteracted at the HPGP firmware level.

The EVExchange~\cite{conti2022evexchange} attack has been proposed to allow an attacker to charge their vehicle while the victim pays.
The authors claim that their attack can operate fully without modifying packets, leaving even TLS connections vulnerable.
We argue that unlike the testbed evaluation conducted in the paper, additional real-world challenges would complicate the attack without MitM access to the communication.
A state transition in basic signaling is synchronized with a transition of the protocol state machine, and measurements of the current should be compared to the reported values.
If messages are passively forwarded as the paper describes, a competent implementation should notice and trigger an emergency stop.
Therefore, enabling TLS would defend against the attack in the real world.

\subsection{Protocol Improvements}
Forward looking researchers are proposing improvements to ISO 15118 standards and the PKI system, such as enabling multiple users to share a PnC enabled vehicle~\cite{plappert2023secure}.
Others have discussed using HSMs to store and generate the PnC credentials~\cite{fuchs2020trust, fuchs2020hip}.
Proposals have also been made to replace PKI~\cite{kailus2024self} with Self-Sovereign Identities, and to transition to quantum safe cryptography~\cite{kern2023quantumcharge}.

\section{Conclusion}

In this paper, we conducted the first and largest real-world measurement study of the EV charging infrastructure, measuring 325 unique charging stations. 
According to our results, the majority of the current charging infrastructure (88\% of tested devices) does not implement TLS and modern parts of the ISO 15118 standard.
While support for the decade-old ISO 15118-2 is slowly rising in new installations, there is no known public deployment of the latest and most secure ISO 15118-20 standard.
We presented results about other implementation details of CCS, and identified vulnerable behavior in the NMK selection of some devices, which lowers the bar for attackers trying to join the HPGP network.
Based on our study, we recommend swiftly deploying TLS, and making a plan to deprecate insecure connections.

In the discussion, the paper presented an overview of the security benefits of TLS, as well as the implementation costs.
Our analysis of the TLS ecosystem compared to the use of PKI on the web highlighted an important and previously unidentified design flaw.
To combat this, we proposed three countermeasures, designed with backwards compatibility and privacy in mind, to detect and limit the impact of a single compromised certificate.
Finally, we discussed the need for improvements to the SLAC process.

\section*{Acknowledgments}

We would like to thank armasuisse Science + Technology for their support. 
Marcell was funded by the Engineering and Physical Sciences Research Council (EPSRC) and Sebastian was supported by the Royal Academy of Engineering and the Office of the Chief Science Adviser for National Security under the UK Intelligence Community Postdoctoral Research Fellowships programme.

\section*{Ethics Considerations}

For both practical and ethical reasons, we modeled our research after commonly conducted Internet-wide scans.
As such, our experiments perform non-malicious CCS charging sessions and collect data in the process.
All tested chargers were installed for public use and are part of critical infrastructure, so we did not want to risk damaging them, or compromising any user data.
We analyze the ethics of our work from the below angles:

\textbf{Data Privacy:}
All of our data was collected using our EV emulator and public chargers.
We did not collect data on the charging sessions of real users charging their car, and there was no foreseeable way in which data from a previous charging session could be leaked to us.
Since the tested devices were provided and owned by companies, none of the information we collected about them could be classified as personal information.

\textbf{Infrastructure Damage:}
Our experiments mostly followed standard-compliant behavior, but deviated in minor non-malicious ways, such as by attempting a connection with a TLS 1.1 client instead of the required 1.2.
Therefore, any CCS implementation should handle our experiments gracefully and we are confident that we caused no damage to tested devices.

\textbf{Public Safety:}
We believe that documenting the current state of security in EV charging is important and necessary.
Our results will inform researchers, companies and legislators, allowing them to push for changes in the industry that will benefit everyone.
In cases where implementations used older, less secure versions of CCS, we treated this as a design choice by manufacturers, which does not require disclosure.
However, in cases where we discovered actual flaws in the implementation, we disclosed them to the affected parties.

\textbf{Data Availability:}
In the paper, we discuss the statistics seen in our dataset, and the companies that implemented TLS.
The dataset is made publicly available consisting of a list of each specific charger we tested, including location, serial numbers, time of test, and key findings.
During our evaluation we found certain issues and potential vulnerabilities which are discussed anonymously.
Our raw data files contain information about these, as well as more sensitive data such as TLS certificates.
Therefore, this detailed data is only made available upon request.

\section*{Open Science Policy}

The code and hardware instructions for the measurement system, basic signaling PCB design files, the dataset, and the data analysis code are available on Zenodo \cite{szakaly_2025_14712107} at \url{https://zenodo.org/records/14712107}.
The dataset includes information about all tested chargers, including location, manufacturer and test results.
Additionally, the measurement code is also available at \url{https://github.com/ssloxford/current-affairs} to support ongoing development.
Our C Python module extension to \texttt{open-plc-utils} is included in the Zenodo archive, as well as at \url{https://github.com/ssloxford/open-plc-utils}.

\bibliographystyle{plain}
\bibliography{references}

\appendix

\section{EV Survey}

\label{app:ev}

The security of a real-world charging session will be determined jointly by the car and chargers capabilities.
As such, it is also important to understand the protocol support in current vehicles.
Gaining access to vehicles for testing is significantly more challenging than gaining access to chargers.
Therefore, to understand the current state of the market we utilized an online survey of public information.

Car manufacturers often publish information about their vehicles, particularly support for ISO 15118-2 Plug and Charge, and ISO 15115-20.
Public databases such as~\cite{evdatabase} exist to easily look up this information, and we present our findings in the results section.

However, this online survey leaves important questions unanswered.
Vehicles which do not advertise support for PnC could still implement it, or they could implement a much simpler ISO 15118-2 TLS client without PnC.
This implementation requires only a list of trusted root certificates, and enabling TLS has clear security benefits.
We study this possibility by additional experiments on real cars, using a charger emulator we developed.

Support for TLS is only possible if it is implemented by both sides of the communication. Therefore, it is important to understand the state of deployment in EVs.
Additionally, unlike chargers, not all EVs are constantly connected to a back-end system, which can make software updates harder to deploy.

\begin{figure}[t]
\centering
\includegraphics[width=0.95\linewidth]{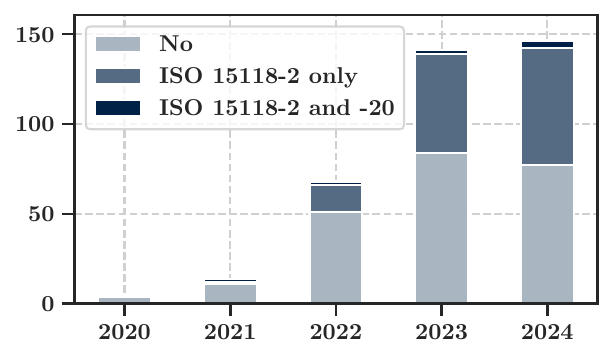}
\centering
\caption{Number of EV models supporting Plug and Charge, by year of first availability. Data from~\cite{evdatabase}. 2024 data includes upcoming announced EVs.}
\label{fig:ev_support}
\end{figure}

Based on data from~\cite{evdatabase}, we found that about half of newly announced cars support Plug\&Charge (PnC) using ISO 15118-2, but only 4 out of 146 EV models scheduled for release in 2024 support ISO 15118-20 PnC.
We plot this trend in Figure~\ref{fig:ev_support}.
It is important to note that these results are not weighted for the sales of each model, the breakdown considers the year a vehicle was first released, and that these features are introduced first into premium vehicles.
Due to these factors, real world adoption of these new standards comes from new or premium vehicles, significantly reducing their current real world market share compared to our data.

\section{Additional Observations}

In this section, we present results that do not have clear security implications, but are interesting and may be useful for future researchers and implementers to better understand the details of real-world CCS deployments.

\subsection{MAC Addresses}

We collected the MAC addresses of the chargers, and analyzed their Organizationally Unique Identifier (OUI) to gain insights into the manufacturers of the embedded systems.
In most systems, MAC addresses can be overwritten by the software or firmware, so our results are merely indicative.
Table~\ref{tab:mac} shows the statistics for the observed OUI fields.

Our results can be classified into three different behaviors between devices.
Some devices were set to a MAC address belonging to an EV charger specific manufacturer, some devices were set to a MAC corresponding to a large generic electronics supplier, and some had special addresses.
Most special addresses had the locally administered bit set, meaning that the address does not need to be unique.
We re-tested some of these chargers a few days apart, and found that they had changed addresses.
For one charger, we also observed it using an address from an IANA region, used for special purpose addresses.

We also noticed that the manufacturer using the NXP Semiconductors OUI set the same, human chosen pattern for the remaining bytes of the address.

\begin{table}
    \centering
    \caption{HLE MAC Addresses of the charger, grouped by the Organizationally Unique Identifier region. We further indicate generic electronics manufacturers, charger manufacturers, and ranges with special purposes.}
    \begin{tabular}{lcr}
        \toprule
        OUI & Devices & Type\\
        \midrule
        Atheros Communications & 1 & Generic\\
        congatec GmbH & 2 & Generic\\
        EcoG & 6 & Charger\\
        GloQuad & 2 & Charger\\
        Huawei Technologies Co.,Ltd & 1 & Generic\\
        I2SE GmbH & 34 & Generic\\
        ICANN, IANA Department & 1 & Special\\
        Kempower Oyj & 10 & Charger\\
        KSE GmbH & 4 & Generic\\
        Locally Administered & 108 & Special\\
        Microchip Technology Inc. & 138 & Generic\\
        NXP Semiconductors & 7 & Generic\\
        Tesla,Inc. & 48 & Charger\\
        Tritium Pty Ltd & 20 & Charger\\
        Unknown & 4 & Special\\
        Wall Box Chargers, S.L. & 1 & Charger\\
        \bottomrule
    \end{tabular}
    \label{tab:mac}
\end{table}

\subsection{IP Addresses}

We collected the IPv6 address given by the charger during the SDP process.
As per the standard~\cite{std_ISO_15118_2}[V2G2-051], link local IPv6 addresses should be used, and all devices should generate it from their MAC address following RFC 4291.
We validate that most devices follow this method, however some devices chose different seemingly random link local addresses.
Because of the SDP process, no specification compliant vehicle should require the IP address to be generated in this way, so we do not expect this to be an issue.
Being able to predict the IP address could be an important step in certain MitM spoofing attacks.

\subsection{Server Ports}

The V2G communication happens using a TCP socket.
The server port is chosen by the charger, and send to the vehicle in the SDP response.
We analyzed the choice of server port, as it reveals information about the internal implementation.

Based on our observations, we were able to classify devices into three categories based on server port selection.
We found devices with ephemeral random, ephemeral incrementing, and constant ports.
By comparing the constants between identical charger models, we can further divide constant ports which appear to be hard coded, and ports which appear to be chosen randomly at system startup.

While there is no clear security implication of this choice based on known attacks, a predictable server port could assist with potential MitM and spoofing attacks.

\balance
\subsection{Attenuation}

\begin{figure}[t]
  \centering
  \includegraphics[width=0.95\linewidth]{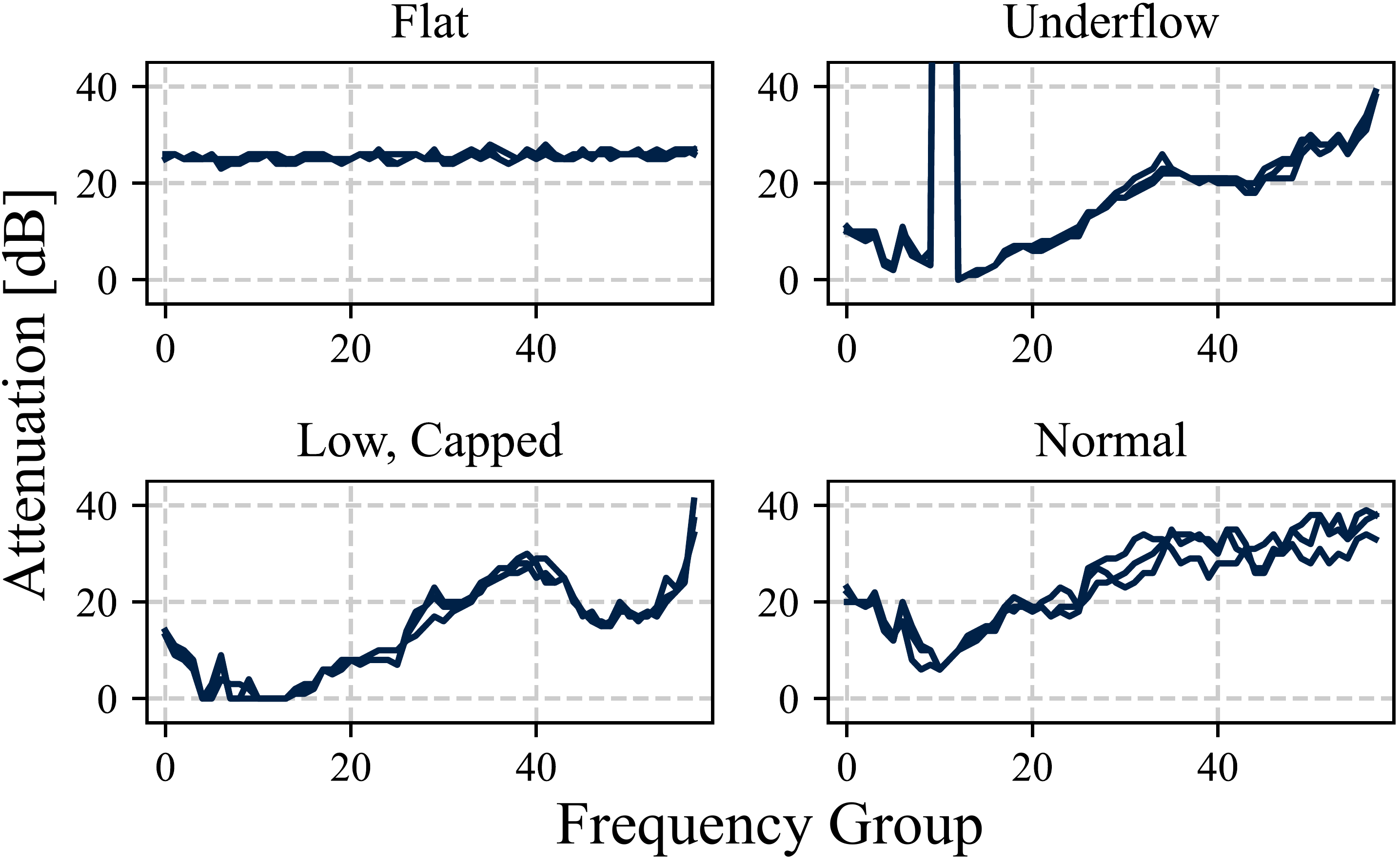}
  \centering
  \caption{Plot of various attenuation profiles received.}
  \label{fig:attenuation_plot}
\end{figure}

During the SLAC process, the charger measures the signal strength of the HPGP packets from the vehicle, and sends this information to it.
Due to the RF nature of the HPGP protocol, this information is provided as a function of frequency.

We observed that all devices returned different attenuation profiles on subsequent measurements, indicating that they do not have a hard coded value, and instead perform a real measurement.
However, we observed three anomalies, and we plot the corresponding signals in Figure~\ref{fig:attenuation_plot}.

First, we observed the shape of the attenuation profile.
Most chargers respond with a similar shape, showing higher attenuation at low frequencies due to the high-pass filter separating the PLC signal from the basic signaling PWM signal, as well as higher attenuation at high frequencies due to various losses.
The difference between min and max attenuation is often in the range of \SIrange{10}{40}{\dB}.
However, one charging manufacturer responds with a very flat profile, within $\pm$\SI{1}{\dB} of the average.
Given that the measurements still contain noise, which varies for each subsequent run, we argue that these are also based on real measurements, but they are likely calculated after some form of channel equalization.

Secondly, different charger models have different attenuation levels.
This could be an issue when two such chargers are installed in close proximity, since a vehicle might accidentally connect to the wirelessly coupled, low attenuation charger as opposed to the wired high attenuation one.
Discrepancies may be due to variations in PLC chips, internal attenuation or charging cables.
In our findings, we see differences of about \SI{10}{\dB} to be common.

Finally, we occasionally observed an underflow in the data.
As per the specification, the attenuation for each frequency is an unsigned byte, however one charger manufacturer would regularly underflow, and send very high (\SI{255}{dB}) attenuation.
It is possible that an implementation averages all the unsigned bytes to determine the total attenuation.
Having a few of these very high values in the average can easily cause the vehicle to calculate a higher average attenuation for the charger it is directly connected to, then for a charger that it is not connected to, and which therefore does not underflow.

We believe that charging station manufacturers should work to calibrate their signal strength measurements to provide consistent, and real results.
The calibration should be done end-to-end, to compensate for the cable, analog front end and PLC chip.
Furthermore, steps should be taken to avoid underflows when reporting the result, as we saw some manufacturers correctly cap their measurements at 0.

From a security perspective, inconsistent measurements during the SLAC process can open the door for a wireless attacker to interfere with the process, and hijack the session.

\end{document}